\documentclass[aps,prl,reprint,superscriptaddress] {revtex4-2}
\usepackage{graphicx}
\usepackage{subfloat}

\begin{document}
\title{Suppressing electron disorder-induced heating of \\ ultracold neutral plasma via optical lattice}

\author{HaiBo Wang}
\affiliation{Hefei National Research Center for Physical Sciences at the Microscale and \\Department of Modern Physics, University of Science and Technology of China, Hefei, Anhui 230026, China}
\affiliation{Hefei National Laboratory, University of Science and Technology of China, Hefei, Anhui 230088, China}

\author{Zonglin Yao}
\affiliation{Hefei National Research Center for Physical Sciences at the Microscale and \\Department of Modern Physics, University of Science and Technology of China, Hefei, Anhui 230026, China}
\affiliation{Hefei National Laboratory, University of Science and Technology of China, Hefei, Anhui 230088, China}

\author{ Fuyang Zhou}
\affiliation{Institute of Applied Physics and Computational Mathematics, Beijing 100088, China}

\author{Yong Wu}
\email[]{wu$_$yong@iapcm.ac.cn}
\affiliation{Institute of Applied Physics and Computational Mathematics, Beijing 100088, China}

\author{Jianguo Wang}
\affiliation{Institute of Applied Physics and Computational Mathematics, Beijing 100088, China}

\author{Xiangjun Chen}
\email[]{xjun@ustc.edu.cn}
\affiliation{Hefei National Research Center for Physical Sciences at the Microscale and \\Department of Modern Physics, University of Science and Technology of China, Hefei, Anhui 230026, China}
\affiliation{Hefei National Laboratory, University of Science and Technology of China, Hefei, Anhui 230088, China}

\date{\today}

\begin{abstract}
Disorder-induced heating (DIH) prevents ultracold neutral plasma into electron strong coupling regime. Here we propose a scheme to suppress electronic DIH via optical lattice. We simulate the evolution dynamics of ultracold neutral plasma constrained by three-dimensional optical lattice using classical molecular dynamics method. The results show that for experimentally achievable condition, electronic DIH is suppressed by a factor of 1.3, and  the Coulomb coupling strength can reach to 0.8 which is approaching the strong coupling regime. Suppressing electronic DIH via optical lattice may pave a way for the research of electronic strongly coupled plasma.
\end{abstract}

\maketitle
Ultracold neutral plasma (UNP) provides an excellent platform for investigating strongly coupled and high-energy-density plasma (HEDP) \cite{1,2,3}. Experimentally, UNPs can be produced by direct photoionization of laser-cooled atoms \cite{4, 5,6,7}, a Bose-Einstein condensate \cite{8}, or from evolution of atomic \cite{PhysRevLett.85.4466}, molecular  Rydberg gases \cite{9}. The extremely low temperature of UNPs enables them to be in or near the strongly coupled regime, and dilute character implies long evolution timescale which ensures them to be precisely diagnosed \cite{5,6,11,12,13}. Unique nature of UNP has also inspired recent experimental investigations on magnetized plasmas \cite{14,15,16,17}, relaxation processes \cite{18,19,20,21}, as well as collective modes \cite{6,23,24,25}, while a series of theoretical works \cite{26,27,28,29,30,31} explored the physical processes in UNPs intensively. Moreover, UNPs are potentially advantageous as a source of high-brightness ion beam \cite{32,33,34,35,36}  for nanoscale measurement and nanofabrication, and a source of high-coherence electron beam \cite{37,38,39}  for coherent diffractive imaging and microscopy. From the plasma physics perspective, one of the main motivations for studying UNPs is the fact that they can be strongly coupled, and have the possibility to create Wigner crystallization \cite{40}. \\

However, the Coulomb coupling strength is limited mainly by disorder-induced heating (DIH), since the plasma is created in a completely uncorrelated state, Coulomb interaction potential energy is rapidly converted into kinetic energy and hence both the electron and the ion components are heated \cite{26,40}. The Coulomb coupling strength $\Gamma  _{\alpha} =e^{2} /\left ( 4\pi\varepsilon _{0} a_{\alpha } k _{B} T_{\alpha }  \right )$, where $e$ represents the elementary charge, $a_{\alpha } =\left ( 3/4\pi n_{\alpha }  \right ) ^{1/3}  $ is the Wigner-Seitz radius for density,  $\varepsilon _{0}$ is the vacuum permittivity,  $k_{B}$ is the Boltzmann constant, and $T_{\alpha }$ denotes the corresponding ion or electron temperature, while $n_{\alpha}$ is the number density of ions or electrons. Therefore, suppressing DIH and improving electronic and ionic Coulomb coupling strength are of great significance for research in UNPs and HEDPs.\\

For the ionic component of UNPs, strong coupling is easier to be achieved \cite{2,45}. Numerous schemes have been proposed to suppress the DIH of ions based on pre-correlating the system before ionization by using a degenerate Fermi gas \cite{26}, Rydberg blockade \cite{43,44}, an optical lattice \cite{42}, or Penning ionization of a molecular Rydberg gas \cite{66}. Among these schemes, Rydberg blockade  \cite{44} and Penning ionization of a molecular Rydberg gas  \cite{66} have shown promising results. Moreover, it has been recognized that the coupling of ion plasma can be promoted by laser cooling the ions immediately after plasma formation \cite{40,41}, and recent experiment of Langin $et$ $al.$ demonstrated laser cooling of ions and achieved a value of  $\Gamma  _{i}$  as high as 11 \cite{46}. \\

Compared to the strong coupling of ions, there are limited works on pursuing strong coupling of electrons. Several mechanisms contribute to the electron heating, such as DIH, three-body recombination (TBR) and Rydberg-electron collisions (REC) \cite{1}. Vanhaecke $et$ $al.$ \cite{49} suggested that adding additional Rydberg atoms to UNP could increase or decrease the electron temperature, and Pohl $et$ $al.$ \cite{50} investigated this idea systematically by simulations and concluded that $\Gamma_{e}$ could be increased to 0.5 by using more sophisticated processes of adding Rydberg atoms. More recent experiment and simulations by Crockett $et$ $al.$ \cite{51} showed that embedding Rydberg atoms into UNP has limited ability to push a plasma into electron strongly coupled regime. Moreover, Tiwari $et$ $al.$ \cite{52} suggested using strong external magnetic field to reduce DIH and TBR of electrons, the basic idea of which is constraining electron motion to reduce electron kinetic energy. This scheme needs a strong magnetic field and can only suppress the heating of electrons in the perpendicular direction to the magnetic field. At the density $n_{e} \simeq 10^{8} \ cm^{-3}$, a strong magnetic field of one-tenth of a Tesla or higher is needed.\\
\begin{figure}
	\includegraphics[width=8cm]{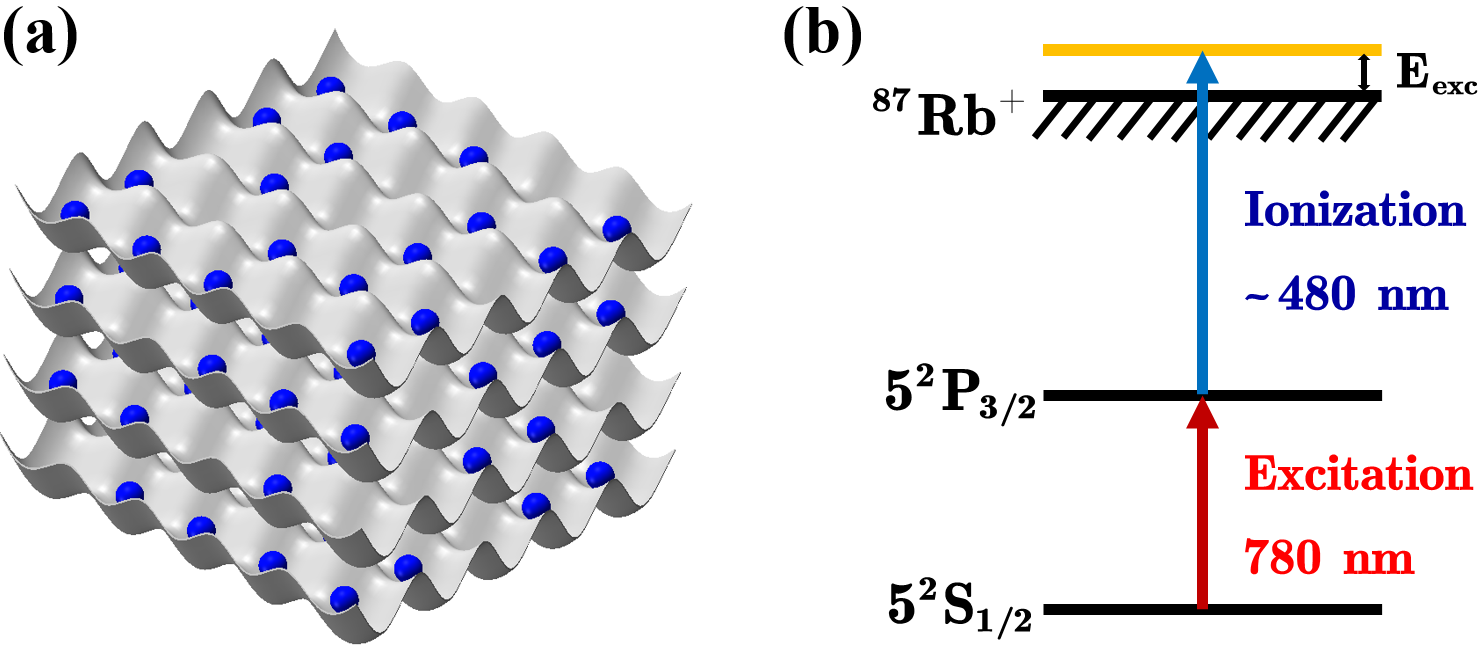}%
	\caption {Schematics of the scheme to suppress DIH. $\left ( a \right )$ Loading atomic cloud into a three-dimensional fractional filling optical lattice, where blue spheres represent cold atoms, and the gray baseboard represents optical trap. $\left ( b \right )$ Two-color photoionization of $^{87}Rb$.}
\end{figure}

In this letter, we propose a scheme to suppress DIH of electrons, and push them into deeper coupling regime. Figure 1 illustrates the schematics of the scheme, taking Rubidium-87 ($^{87} Rb$) as a prototype. The creation of $^{87} Rb$ UNP starts from laser cooled and trapped neutral atoms in a magneto-optical trap (MOT) using $5^{2} S_{1/2}$-$5^{2} P_{3/2}$ transition at 780 nm. Up to $10^{9}$ atoms can be cooled to as low as microkelvin temperatures and confined at densities approaching $10^{11} cm^{-3}$. $5^{2}P_{3/2}$ atoms are photoionized near the ionization threshold by a laser pulse at $\sim $ 480 $nm$. UNPs are then formed having electron temperatures in the 1-1000 kelvin range and ion temperatures from tens of millikelvin to a few kelvin. The typical spherical Gaussian density distribution of atomic cloud, and the threshold photoionization leads to a disordered configuration of ions and electrons. The subsequent rearrangement of ions and electrons establishes interparticle correlations, and thus decreases the potential energy of plasma. The conservation of energy leads to a rapid heating of ions and electrons, i.e. DIH. This DIH prevents UNP into electron strong coupling regime. In order to suppress the DIH, pre-ordering ions and electrons is a natural consideration. As shown in Fig. 1$\left ( a \right ) $, the cooled atoms are first loaded into a three-dimensional optical lattice to establish the pre-ordering of atoms. The ionic component of the subsequently-created UNP will inherit the ordering of atoms and constrain the motion of electrons. As a result, the DIH of electrons will be suppressed.\\

A classical molecular dynamics (MD) \cite{67} simulation is performed to simulate the evolution dynamics of the optical lattice constrained UNPs. The present work focuses on the DIH of electrons, which mainly occurs at early phase of timescale of 1-2  $\tau _{e} \left ( \sim 1 ns \right )$ , where $\tau _{e}$  is equal to the inverse electron plasma frequency, i.e. $\tau _{e} =\omega _{pe}^{-1}=\sqrt{m_{e}\varepsilon _{0}/\left ( n_{e}e^{2} \right ) }$ . Thus, the simulation runs for $t_{max}\omega _{pe}=10$ scaled time units. On the other hand, the characteristic plasma expansion time is determined by $ \tau _{exp}=\sqrt{m_{i}\sigma \left ( 0 \right )^{2}/k_{B}\left [ T_{e}\left ( 0 \right ) + T_{i}\left ( 0 \right )\right ] } $, typically 1 $\mu s$, where  $\sigma \left ( 0 \right )$ is the initial size of the plasma cloud and $T_{e}\left ( 0 \right )$ , $T_{i}\left ( 0 \right )$ are the initial electron and ion temperatures, respectively \cite{60}. Therefore, under the present simulation timespan, the plasma hardly expands. The radius of the simulation spherical volume is set at 0.5 mm which is large enough to model the dynamics of UNPs and avoid sample size effect. The UNP is put at the center of the spherical volume. In addition, the timestep is set at a few femtoseconds that is sufficiently short to allow us revealing the ‘real’ evolution of UNPs.\\

The Coulomb interaction potential of all charged particle pairs is adopted in the simulations. In order to avoid singularities, the Coulomb potential is given by the form \cite{28}
\begin{equation}
    \pm \frac{e^{2}}{4\pi \varepsilon_{0}\sqrt{r^{2}+\left ( \eta a_{\alpha }  \right )^{2}  }  }
\end{equation}
where $r$ is the separation between two charged particles,  $\eta$ is an adjustable parameter, and the sign $ +  \left ( -  \right ) $ represents Coulomb repulsion (attraction) force, respectively. The choose of $\eta$ is critical to the dynamics of UNP. Tiwari $et$ $al.$ \cite{52} showed that the electron temperature evolution behavior does not appreciably change for $\eta <0.01$ , and thus $\eta $ is chosen to be 0.01. Moreover, TBR and REC are not considered in the simulations because TBR and REC should not significantly increase the electron temperature at early evolution phase as showed by Lyon $et$ $al.$ \cite{63}. We take a kinematic definition of temperature as the average kinetic energy per particle of type $\alpha$ , and defined as
\begin{equation}
    T_{\alpha } =\frac{1}{3Nk_{B}\sum_{i=1}^{N}m_{\alpha } \left | \vec{v} _{\alpha ,i} \right | ^{2} }
\end{equation}
where $N$ is the total number of electrons or ions. At the early phase of plasma evolution, electrons are not in thermal equilibrium state and the distribution of electron velocity is not Maxwellian. The electron temperature is not thermodynamic, but rather interpreted in terms of the kinetic temperature of Eq. (2). \\

As the typical plasma density is $ 10^{8}$-$10^{11} cm^{-3}$, we perform the simulations at different average densities of $10^{8} cm^{-3}$, $10^{9} cm^{-3}$, $10^{10} cm^{-3}$ and $10^{11} cm^{-3}$. The ions are initialized at optical lattice sites for different lattice types such as simple cubic (sc), body-centered cubic (bcc), and face-centered cubic (fcc), which can experimentally be realized with suitable laser arrangements \cite{53,54,55}. For saving the computational cost, we choose $\sim$ 1000 ions in the simulations. This number may differ slightly for different lattice types. Combined with average density and ion number, the initial volume can be determined. Experimentally, lattice spacing is determined by the laser wavelength and arrangement. Therefore, in the present simulations, filling fraction is quite low which corresponds to low density UNP. Due to the neutral character of UNP, the number of electrons equals to ions. The initial electron positions are randomly set. In order to give maximum correlations to develop, the initial ion and electron temperatures (or more precisely, kinetic energies) are set to zero, corresponding to $\Gamma _{e}\left ( t=0 \right ) =\infty$  and  $\Gamma _{i}\left ( t=0 \right ) =\infty$ \cite{28}. For $^{87}Rb$ UNP,  $m_{i}/m_{e}$ is equal to 1.65. \\
\begin{figure}
 \includegraphics[width=8.5cm]{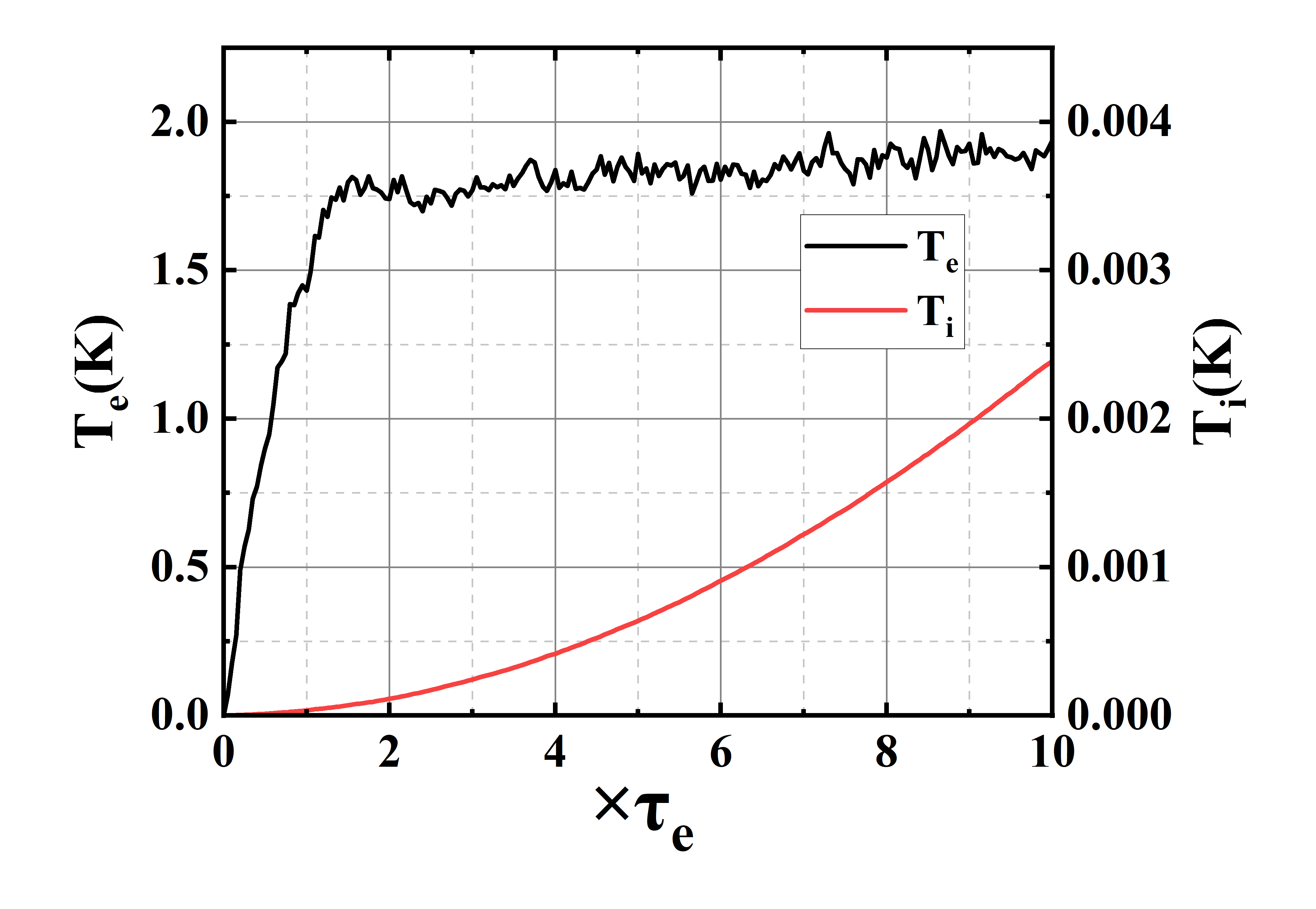}%
 \caption {Electron and ion temperature evolution on timescale of 0-10 $\tau _{e}$. Simulations were carried out with initial temperature $T_{e}=T_{i}=0$ and a uniform random distribution of electrons and ions with the average density $n_{e}=n_{i}=10^{8} cm^{-3}$.}
 \end{figure}

Furthermore, we evaluate the influence of optical lattice laser on electron temperature. The optical dipole potential arises from the interaction of the induced atomic dipole moment with the intensity gradient of the laser field \cite{56}. For a Gaussian laser beam, the typical trap depth can be up to millikelvin  \cite{53}. For $^{87}Rb^{+}$, the induced ionic dipole moment with the dipole laser is tremendously small so the interaction potential of induced ionic dipole moment with plasma electrons is relatively weak, compared to the coulomb potential between plasma ions and electrons. Therefore, the influence of optical lattice laser is negligible.\\

Firstly, we simulate the evolution of normal UNPs where ions are randomly distributed. For the normal UNPs, the DIH of electrons occurs on a timescale of 1-2 $\tau _{e}\left ( \sim 1ns \right )$. Then, the electrons undergo further heating by TBR and REC on a slower timescale, which is due to the formation of highly excited Rydberg atoms and the transfer of Rydberg atoms to more deeply bound levels. In addition, adiabatic cooling accomplished by plasma expansion will decrease electron temperature on a timescale of about 10 $ \mu s$. Our simulation time spans the timescale of 10 $\tau _{e}$, and enables us to explore the main behavior of electronic DIH. Compared to the electronic DIH, the ionic DIH happens on a rather slower timescale of about 1 $\mu s$ due to large ion mass.\\

\begin{figure}
	\centering
\includegraphics[width=7cm]{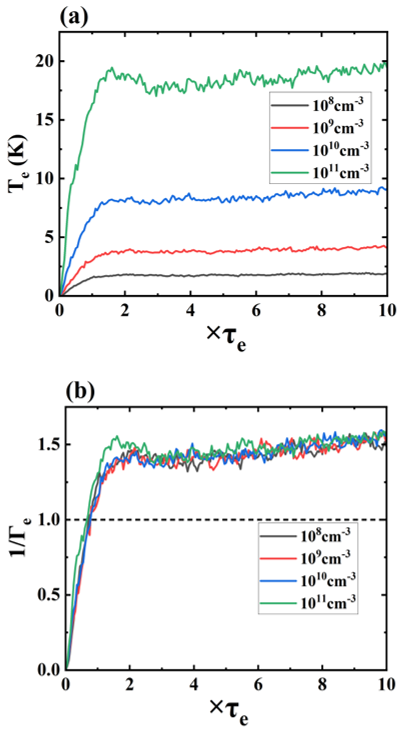}%
\caption{ Electron temperature $\left ( a \right )$ and Coulomb coupling strength $\left ( b \right )$ evolution at different initial average density:$10^{8} cm^{-3}$ $\left (black \right )$, $10^{9} cm^{-3}$ $\left (red \right )$, $10^{10} cm^{-3}$ $\left (blue \right )$ and $10^{11} cm^{-3}$ $\left (green \right )$. }
\end{figure}

Figure 2 presents the temporal evolution of average kinetic temperature of electrons $\left ( black \ line \right )$ and ions $\left ( red \  line \right )$. One can find that the electron temperature rapidly increases and approaches to a plateau within nanosecond showing the obvious behavior of the DIH of electrons. The result agrees well with previous experiments \cite{5,13} and simulations \cite{57,52}. During the process of DIH, electrons gain kinetic energy largely via the ballistic motion associated with Coulomb attraction to the ions \cite{52}. The typical ion equilibration timescale $\tau _{i}$ is about 1 $\mu s$, which is equal to the inverse ion plasma frequency $\omega _{pi}^{-1}=\sqrt{m_{i}\varepsilon _{0}/\left ( n_{i}e^{2} \right ) }$. In the present simulation timespan, as expected, ions are far from equilibration.\\

 \begin{figure*}
	\centering
	\includegraphics[width=14cm]{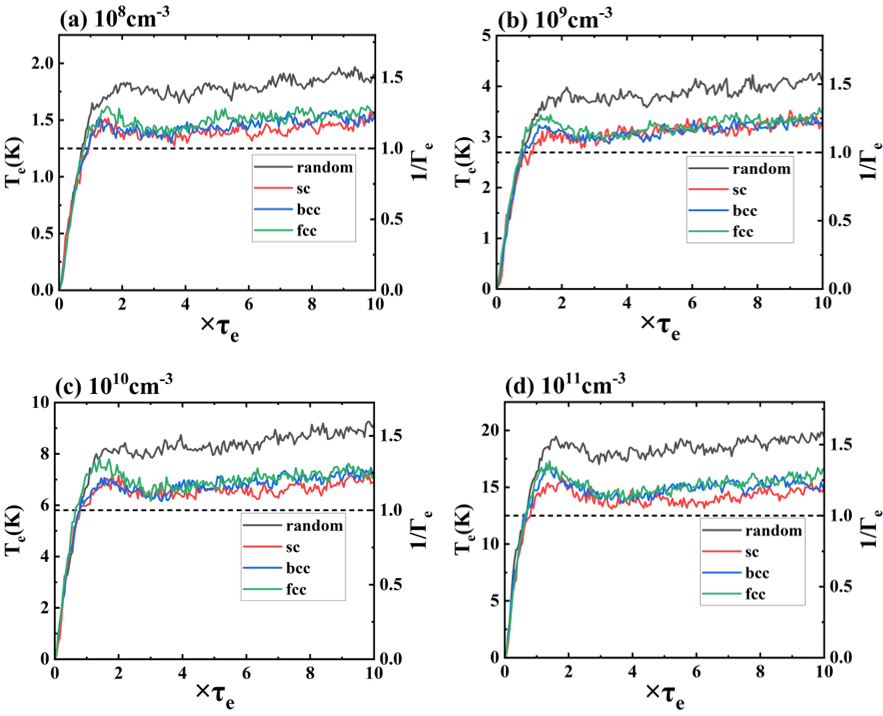}%
	\caption{ Temperature and Coulomb coupling strength evolution of electrons at different cubic lattice geometries and densities: $\left ( a \right )$ $10^{8} cm^{-3}$, $\left ( b \right )$ $10^{9} cm^{-3}$, $\left (c \right )$ $10^{10} cm^{-3}$, $\left ( d \right )$ $10^{11} cm^{-3}$. The legend represents initial different ion density distributions. }
\end{figure*}

The influence of plasma density on electronic DIH and Coulomb coupling parameter is also investigated. The results are shown in Fig.3 $\left ( a \right )$  and $\left ( b\right )$ . As can be seen, with the increase of density, the plateau temperature $T_{DIH}$ increases. This can be easily understood because the electron kinematic energy gains from the Coulomb interaction $k_{B}T_{DIH }\sim e^{2} /\left (4\pi \varepsilon _{0}a_{e}  \right ) \propto n_{e}^{1/3}$ . It is interesting to note that, although plateau temperature $T_{DIH}$ increases with the plasma density, the Coulomb coupling strength keeps unchanged, as shown in Fig. 3 $\left ( b \right )$. According to the definition of Coulomb coupling strength $ \Gamma  _{e} =e^{2} /\left ( 4\pi\varepsilon _{0} a_{e } k _{B} T_{e }  \right ) $, which represents the ratio of Coulomb interaction energy to electron kinematic energy, we have $k_{B}T_{DIH} \sim e^{2}/\left ( 4\pi \varepsilon _{0} a_{e} \right ) $. Therefore  $\Gamma  _{e}$ approaches to a constant value regardless of the initial plasma density. \\

Gericke $et$ $al.$ \cite{42} suggested confining atoms into an optical lattice to suppress the DIH of ions. After that, by using classical MD methods, several theoretical works \cite{57,58,61} demonstrated that the DIH of ions can indeed be suppressed. In the present work, we focus on possibility of suppression of the DIH of electrons by loading atoms into a three-dimensional optical lattice. The pre-ordering is obtained by setting ions into lattice sites. Figure 4 presents electron temperature and Coulomb coupling strength evolution for different lattice types and plasma densities. It is quite straightforward that, in all cases, the electronic DIH is significantly suppressed. The plateau temperature is reduced by a factor of 1.3. Meanwhile, the electronic Coulomb coupling strength increases to 0.8, approaching to strong coupling regime. One can also find that electron plateau temperature increases with the increasing density, while the electronic Coulomb coupling strength keeps unchanged which is similar to Fig. 3. For different types of lattice, the electron number density differs slightly. Therefore, the plateau temperature as well as the electronic Coulomb coupling strength, are almost independent on lattice types. \\

As we have mentioned that Tiwari $et$ $al.$  \cite{52} suggested using one dimensional strong external magnetic field to constrain electron motion to reduce electron kinetic energy and thus to cool electrons. This scheme needs strong magnetic field and can only suppress the heating of electrons in the direction perpendicular to the magnetic field. Furthermore, due to large ion mass, the motions of ions can hardly be constrained by the external magnetic field, and thus the DIH of ions could not be extenuated. Previous theoretical works \cite{57,58,61} have demonstrated that optical lattices can push the ionic components into deeper coupling regime. Therefore, based on our simulations,both the DIH of electrons and ions can be suppressed via optical lattice. \\

In conclusion, we propose a scheme to suppress the DIH of electrons, which is based on pre-ordering UNPs by loading atoms into three-dimensional optical lattices. By performing the classical MD simulations, we demonstrate that the DIH of electrons could be suppressed by a factor of 1.3, compared to the conventional UNPs. We also show that the degree of electron temperature reduced is independent on plasma density and lattice types. Both electronic and ionic Coulomb coupling strength can be pushed into deeper coupling regime, which will allow us for exploring novel and intriguing phenomena. Moreover, decreased electron and ion temperature will improve the coherent length of electron beam and brightness of focused ion beam based on UNPs.\\

\begin{acknowledgments}
This work is supported by Innovation Program for Quantum Science and Technology (Grant No.2021ZD0303300), the National Key Research and Development Program of China (Grant No. 2022YFA1602502) and the National Natural Science Foundation of China (Grant No.12127804).
\end{acknowledgments}

\bibliography{refs.bib}

\end{document}